\documentclass[a4paper,11pt]{article}
\usepackage{jheppub} 
\usepackage{amsmath}
\usepackage{lineno}
\usepackage{graphicx}
\usepackage{xcolor}
\usepackage{physics}
\usepackage{appendix}
\usepackage{tcolorbox}
\usepackage{bm}

\newcommand{\be}{\begin{eqnarray} }
\newcommand{\ee}{\end{eqnarray} }
\newcommand{\bs}{\begin{split} }
\newcommand{\es}{\end{split} }

\title{Cosmological cutting rules for Bogoliubov initial states: any mass and spin}

\author[a,1]{Diptimoy Ghosh}
\affiliation[a]{Indian Institute of Science Education and Research Pune \\Pune 411008, India}

\author[a,2]{Farman Ullah}
\affiliation[]{}
\emailAdd{diptimoy.ghosh@iiserpune.ac.in}

\emailAdd{farman.ullah@students.iiserpune.ac.in}

\abstract{The cosmological optical theorem and the cutting rules are well-known consequences of unitary time evolution in cosmology. The earlier works showed that assuming a Bunch-Davies initial state, one can derive equations relating a wavefunction diagram with a given number of internal lines to a sum of diagrams with fewer internal lines. In particular, it can relate a loop diagram to a sum of tree-level diagrams. Recently these relations were generalised to a set of excited initial states known as Bogoliubov states and they were shown to have non-trivial consequences for $n$-point contact and 4-point exchange diagrams. This analysis restricted the field content to massless and conformally coupled scalar fields. In this paper, we take the final step of generalising these ``Bogoliubov cutting rules" to fields of any mass and spin. We define modified propagator identities and corresponding ``Discontinuities" which automatically generalise the earlier relations to fields of any mass and spin. Finally, we discuss issues concerning the far past convergence of time integrals in the complex plane.}

\begin{document}
\maketitle
\flushbottom

\section{Introduction}
Understanding the implications of unitarity for cosmological observables is a topic which has attracted a lot of attention. In \cite{Goodhew:2020hob,Melville_2021, Goodhew:2021oqg}, a set of unitarity relations was derived for the cosmological wavefunction coefficients collectively called the cosmological cutting rules. These relations were derived assuming a Bunch-Davies initial state which is certainly the simplest and the most natural choice. However, inflation \cite{Starobinsky:1980te,Guth:1980zm,Linde:1981mu} could have started in a different state \cite{Starobinsky:2001kn,Starobinsky:2002rp,Agullo:2010ws,Ganc:2011dy,Holman:2007na,Easther:2002xe,Brandenberger:2002hs,Meerburg:2009ys,Jain:2022uja,Ghosh:2022cny,Ansari:2024pgq,Chopping:2024oiu} and therefore, in \cite{Ghosh:2024aqd} we generalised these cutting rules to Bogoliubov initial states \cite{Allen:1985ux,Shukla:2016bnu}. The mode functions corresponding to these states are complex linear superpositions of Bunch-Davies ones. Therefore, naturally, these states are parametrised by two complex functions $\alpha_k, \beta_k$ known as Bogoliubov coefficients giving rise to an infinite set of excited states. Due to the simple relation between the two sets of mode functions, in \cite{Ghosh:2024aqd} we were able to derive prescriptions for getting $n$-point contact and 4-point exchange  Bogoliubov wavefunction coefficients from Bunch-Davies ones without performing explicit computations. Earlier such a prescription was derived for only contact three-point functions \cite{Jain:2022uja,Ghosh:2023agt}.
  Although the Bunch-Davies cutting rules were generalised to fields of any mass and spin \cite{Melville_2021,Goodhew:2021oqg}, the Bogoliubov cutting rules derived in \cite{Ghosh:2024aqd} are only valid for massless and conformally coupled scalars. Therefore, in this work, we take the final step of \textit{generalising these rules to fields of any mass and spin}. It turns out that it is straightforward to generalise cutting rules to spinning fields once they are generalised to scalar fields of any mass. However, there are two apparent obstructions to extending the rules to massive scalars:
  \begin{itemize}
      \item \textbf{Propagator identities:} In the case of a Bunch-Davies initial state, the propagator identity, $\mathcal{K}^{*}_{-k}=\mathcal{K}_{k}$, is valid for scalar fields of any mass provided we approach the negative energies from below i.e., $-k=e^{-i\pi}k$\footnote{See \cite{Goodhew:2020hob,Melville_2021, Goodhew:2021oqg} for more details.}. However, in the case of a Bogoliubov initial state, there are two propagator identities,
      \begin{align}
           \label{1:id} K_{-k}^*(\alpha^{*}_k,\beta^{*}_k) &=K_{k}(\alpha_k,\beta_k)\,,\\
        \label{2:id} K_{k}^*(\beta^{*}_k,\alpha^{*}_k) &=K_{k}(\alpha_k,\beta_k) \,. 
      \end{align}
     and the identity, $K_{-k}^*(\alpha^{*}_k,\beta^{*}_k) =K_{k}(\alpha_k,\beta_k)$, does not extend to massive scalars. Since the mode function includes both types of Hankel functions (see \eqref{h:2}, \eqref{h:1} \& \eqref{mode}), no matter how one approaches the negative energies (from above or below), the propagator identity mentioned above is never satisfied owing to the transformation properties of the two Hankel functions given in \eqref{H:1} \& \eqref{H:2}. Upon some inspection, it turns out that the massive scalar propagator instead satisfies a modified identity,
      \begin{align}
           K^{*}_{-k}((\alpha_k+2i\beta_k\cos{\pi\nu})^{*},\beta^{*}_k,\tau)=K_{k}(\alpha_k,\beta_k,\tau) \,,
      \end{align}
      where $\nu=\sqrt{\frac{9}{4}-\frac{m^{2}}{H^{2}}}$. Thankfully, this identity reduces to $K_{-k}^*(\alpha^{*}_k,\beta^{*}_k) =K_{k}(\alpha_k,\beta_k)$ in the case of a massless ($\nu=3/2$) or a conformally coupled scalar field ($\nu=1/2$). Therefore, one obtains a generalised identity valid for scalar fields of any mass. The propagator identity, $K_{k}^*(\beta^{*}_k,\alpha^{*}_k) =K_{k}(\alpha_k,\beta_k)$, extends trivially to massive scalars.     
       
      \item \textbf{Analytical continuation:} Since the negative energies are always approached from below, one analytically continues the wavefunction coefficients to the lower half complex plane. In the case of Bunch-Davies initial state, since the mode functions contain only positive frequencies, the analytic continuation to the lower half plane is well defined i.e., the time integrals in the past converge $\forall k_i \in \mathbb{C}^{n-}$. In the case of Bogoliubov initial states, the mode functions have both positive and negative frequency which for the propagator implies,
      \begin{align}
         \lim_{\tau\rightarrow-\infty}  K_k(\alpha_k,\beta_k,\tau) \sim e^{i\text{Re}(k)\tau}e^{-\text{Im}(k)\tau}+ e^{-i\text{Re}(k)\tau}e^{\text{Im}(k)\tau}\,, \hspace{10mm}\text{where} \hspace{1mm} k\in \mathbb{C}\,. 
      \end{align}
     The time integrals will converge neither in the upper nor in the lower plane since one of the two terms will always diverge in the far past. However, since we always employ an adiabatic function to turn off interactions in the past, we will with some care be able to analytically continue the wavefunction to the complex plane. 
  \end{itemize}
\noindent The cutting rules can be practically very useful in understanding loop effects in cosmology while computing only tree diagrams \cite{Melville_2021}. This is in particular because, due to the lack of time translational invariance the computations become extremely challenging (even tree-level diagrams can produce logarithms, unlike in the flat space case).  The loop effects in cosmology have been widely investigated following the seminal papers \cite{Weinberg:2005vy,Senatore:2009cf,Senatore:2012nq,Pimentel:2012tw} for the power spectrum. Recently, attempts have been made to extend these calculations to three-point functions \cite{Bhowmick:2024kld,Bhowmick:new}. One can hope that the cutting rules can help us understand aspects of loop effects during inflation without having to perform difficult bulk integrals, see \cite{Melville_2021} for a discussion.\\ \\
 The rest of this paper is organised as follows. In Sec. \ref{propagator:identities} we list modified propagator relations for fields of any mass and spin. In Sec. \ref{cutting:rule} we give the cutting rule relation valid for fields of any mass and spin. In Sec. \ref{convergence} we discuss issues regarding convergence of time integrals. We conclude in Sec. \ref{conclusion}.
\label{sec:intro}
 \\



\noindent{\bf Notations and Conventions} \vspace{2 mm} \\
The spatially flat cosmological spacetimes are defined via the FRW metric given as,
\begin{align}
    ds^{2}=a^{2}(-d\tau^{2}+d\vec{x}^{2}) \,,
\end{align}
where $a$ is the scale factor and $\tau$ is conformal time. For a flat de Sitter FRW background ($a=-1/H\tau$) we get,
\begin{align}
ds^{2}=\frac{-d\tau^{2}+d\vec{x}^{2}}{\left(H\tau\right)^{2}}   \,, 
\end{align}
where $H$ is the Hubble parameter. The FRW time coordinate is denoted by $t$. Derivatives with respect to conformal time are denoted by primes.\\
The field theoretic wavefunction is parameterised as
\begin{align} \label{wavefunction}
    \Psi[\phi,\tau_0]=\text{exp}\left(+\sum_{n=2}^{\infty} \int \frac{d^{3}k_1 \dots d^{3}k_n}{n!( 2\pi)^{3n}}\psi_{n}\phi_{\vec{k}_1}\dots \phi_{\vec{k}_n} \right) \,,
\end{align}
where $\psi_n=\psi_{n}(\{\vec{k}\},\tau_0)$ are the wavefunction coefficients at time $\tau_0$. The external momenta are denoted by $\{\vec k_i\}$. In the rest of the paper, internal energies are denoted by $\{p_m\}$. By energy we mean the modulus of the momenta, $k_i=|\vec k_i|$. We denote by $\mathcal{K}$ and $\mathcal{G}$ the bulk-boundary and bulk-bulk propagators for a Bunch-Davies initial state and $K$ and $G$ for the case of a Bogoliubov initial state. For useful references on the field-theoretic wavefunction in the cosmological context see \cite{Anninos:2014lwa,Guven:1987bx,Maldacena:2002vr,Ghosh:2014kba,Maldacena:2011nz,Arkani-Hamed:2017fdk,Benincasa:2022gtd,Chakraborty:2023yed}.


\section{Cutting rules for any mass and spin}
In this section, we will extend the Bogoliubov cutting rules to fields of any mass and spin. Since the form of the bulk-bulk propagator as a function of the bulk-boundary propagator is unchanged, one only needs to define appropriate ``Discontinuity" operations (Disc) to extend the cutting rules. The definition of the Disc operation will be determined by the propagator identities discussed below.


\subsection{Propagator identities and corresponding ``Discontinuities"}
\label{propagator:identities}
One of the key ingredients to deriving the cutting rules is propagator identities. In the case of a Bunch-Davies initial state, the identity reads,
\begin{align}
    \mathcal{K}^{*}_{-k}=\mathcal{K}_k \label{BD:identity} \,.
\end{align}
This identity is crucial in identifying the appropriate Disc operation for the cutting rules. It was shown in \cite{Goodhew:2021oqg} that this identity generalises to fields of any mass and spin. In \cite{Ghosh:2024aqd}, we discussed  analogous propagator identities for a Bogoliubov initial state for massless and conformally coupled scalars,
\begin{align}
  \label{disc:1}  K_{-k}^*(\alpha^{*}_k,\beta^{*}_k) &=K_{k}(\alpha_k,\beta_k)\,,\\
    \label{disc:2}K_{k}^*(\beta^{*}_k,\alpha^{*}_k) &=K_{k}(\alpha_k,\beta_k) \,.
\end{align}
We stress that these relations are derived by keeping the $k$ dependence of $\alpha_k,\beta_k$ separate from the rest of the $k$ dependence in $K_{k}(\alpha_k,\beta_k)$. 

\subsubsection{Massive scalars}
To see which of these identities extends to the massive case, we highlight some salient properties of the massive mode functions. It is well known that the mode functions in this case are Hankel functions,
\begin{align}
   \label{h:2} \varphi^{+}_k(\tau)=ie^{-i\frac{\pi}{2}(\nu+\frac{1}{2})}\sqrt{\pi}\frac{H}{2}(-\tau)^{\frac{3}{2}}H^{(2)}_{\nu}(-k\tau)\,,\\
    \label{h:1}  \varphi^{-}_k(\tau)=-ie^{i\frac{\pi}{2}(\nu+\frac{1}{2})}\sqrt{\pi}\frac{H}{2}(-\tau)^{\frac{3}{2}}H^{(1)}_{\nu}(-k\tau)\,,
\end{align}
where $\nu=\sqrt{\frac{9}{4}-\frac{m^{2}}{H^{2}}}$. In the case of a Bunch-Davies initial state only $\varphi^{+}_k(\tau)$ survives. However, for a Bogoliubov initial state, the mode functions multiplying the raising operator, $\Phi^{+}_{k}$, is a linear combination of the two solutions given above,
\begin{align}
   \label{mode} \Phi^{+}_{k}(\alpha_k,\beta_k,\tau)=\alpha_{k}  \varphi^{+}_k(\tau) + \beta_{k}\varphi^{-}_k(\tau)\,.
\end{align}
The corresponding bulk-boundary propagator is given by,
\begin{align}
    K_k(\alpha_k,\beta_k,\tau)=\frac{ \Phi^{+}_{k}(\alpha_k,\beta_k,\tau) }{ \Phi^{+}_{k}(\alpha_k,\beta_k,\tau_0) }\,.
\end{align}
Let us list some useful properties of Hankel functions before we derive the propagator identities. 
\begin{align}
&{H^{(1)*}_\nu}(z^{*})={H}^{(2)}_{\nu^{*}}(z)\,,\\
    &{H}^{(1)}_{-\nu}(z)= e^{i\pi\nu}{H}^{(1)}_{\nu}(z)\,,\\
    & {H}^{(2)}_{-\nu}(z)= e^{-i\pi\nu}{H}^{(2)}_{\nu}(z)\,.
\end{align}
Using above equations one can immediately see that $[\Phi^{+}_{k}(\beta^{*}_k,\alpha^{*}_k,\tau)]^{*}=\Phi^{+}_{k}(\alpha_k,\beta_k,\tau)$ and consequently the propagator identity \eqref{disc:2} is true for massive scalars also. Therefore, we define the corresponding Disc operation as Disc(1) given in \cite{Ghosh:2024aqd},
\begin{align}
    &\underset{\{\alpha_{p_m},\beta_{p_m}\}}{\text{Disc(1)}}[f(\{\alpha_{k_i},\beta_{k_i}\},\{\alpha_{p_m},\beta_{p_m}\},\{k_i\},\{\vec{k}_i\},\{p_m\})]= \nonumber \\ 
& f(\{\alpha_{k_i},\beta_{k_i}\},\{\alpha_{p_m},\beta_{p_m}\},\{k_i\},\{\vec{k}_i\},\{p_m\})  - f^{*}(\{\beta^{*}_{k_i},\alpha^{*}_{k_i}\},\{\alpha_{p_m},\beta_{p_m}\},\{k_i\},\{-\vec{k}_i\},\{p_m\})  \,.
\end{align}
The cutting rules involving Disc(2) are non-trivial to extend because the analytical continuation of the bulk-boundary propagator to negative energies is not as straightforward as in the Bunch-Davies case. We explain this below. \\\\
\noindent  Note that Hankel functions have a branch point at $z=0$. Therefore, we will follow the standard convention of choosing the principal branch i.e., $-\pi<\text{arg}(z)<\pi$ (branch cut on the negative $x$-axis). Approaching the branch cut from above, we get,
\begin{align}
   \label{H:1} H^{(1)}_{\nu}(e^{i\pi}z)&=-e^{-i\pi\nu} H^{(2)}_{\nu}(z) \,, \\
    \label{H:2}  H^{(2)}_{\nu}(e^{i\pi}z)&=e^{i\pi\nu}H^{(1)}_{\nu}(z)+(2\cos{\pi \nu})H^{(2)}_{\nu}(z)\,.
\end{align}
Using the properties mentioned above, it was shown that for the Bunch-Davies case if one approaches the negative $k$-axis from below, the propagator identity \eqref{BD:identity} remains intact. The reason is very simple: the bulk-boundary propagator for the Bunch-Davies case is given by,
\begin{align}
    \mathcal{K}_{k}(\tau)=\frac{\varphi^{+}_{k}(\tau)}{\varphi^{+}_{k}(\tau_0)} \,,
\end{align}
and if one defines $-k=e^{-i\pi}k$ then, using the properties above, it is straightforward to check that $[\varphi^{+}_{-k}(\tau)]^{*}=i{\varphi^{+}_{k}}(\tau)$ and therefore, $\mathcal{K}^{*}_{-k}=\mathcal{K}_k$. Let us now analyse the Bogoliubov case. It is immediately clear from \eqref{H:1} \& \eqref{H:2} that $K_{-k}^*(\alpha^{*}_k,\beta^{*}_k) \neq K_{k}(\alpha_k,\beta_k)$ since, $[\Phi^{+}_{-k}(\alpha^{*}_k,\beta^{*}_k,\tau)]^{*} \neq c \hspace{1pt} \Phi^{+}_{k}(\alpha_k,\beta_k,\tau)$, where $c \in \mathbb{C}$. More precisely, 
\begin{align}
     [\Phi^{+}_{-k}(\alpha^{*}_k,\beta^{*}_k,\tau)]^{*}=i\Phi^{+}_{k}(\alpha_k-2i\beta_k \cos{\pi \nu^{*}},\beta_k,\tau) \label{newdisc} \,.
\end{align}
However, one is still able to define a modified propagator identity as follows,
\begin{align}
      K^{*}_{-k}((\alpha_k+2i\beta_k\cos{\pi\nu^{*}})^{*},\beta^{*}_k,\tau)=K_{k}(\alpha_k,\beta_k,\tau) \,. \label{newidentity}
\end{align}
Notice that in the massless or conformally coupled limit the above expression reduces to $ K^{*}_{-k}(\alpha^{*}_k,\beta^{*}_k,\tau)=K_{k}(\alpha^{*}_k,\beta^{*}_k,\tau)$ as expected. Motivated by \eqref{newidentity}, we can define a new disc operation (we call it \text{Disc}($\tilde{2}$)) as follows,
\begin{align} 
&\underset{\{\alpha_{p_m},\beta_{p_m}\}}{\text{Disc($\tilde{2}$)}}[f(\{\alpha_{k_i},\beta_{k_i}\},\{\alpha_{p_m},\beta_{p_m}\},\{k_i\},\{\vec{k}_i\},\{p_m\})]= \nonumber \\ 
& f(\{\alpha_{k_i},\beta_{k_i}\},\{\alpha_{p_m},\beta_{p_m}\},\{k_i\},\{\vec{k}_i\},\{p_m\}) -\nonumber \\
& f^{*}(\{\alpha^{*}_{k_i}-2i\beta^{*}_{k_{i}} \cos{\pi\nu},\beta^{*}_{k_i}\},\{\alpha_{p_m},\beta_{p_m}\},\{-k_i\},\{-\vec{k}_i\},\{p_m\}) \,. 
\end{align}
This generalises the cutting rules involving Disc(2) to massive scalars. As expected for $\nu=3/2$ \& 1/2 this Disc operation reduces to the Disc(2) operation for massless and conformally coupled scalars mentioned in \cite{Ghosh:2024aqd}. Again, we stress that negative energies are approached from below i.e., $-k = e^{-i\pi}k$. To summarize, we give the two propagator identities for massive scalars,
\begin{align}
   \label {id:1} K_{k}^*(\beta^{*}_k,\alpha^{*}_k)& =K_{k}(\alpha_k,\beta_k)  \,, \\
    \label {id:2}  K^{*}_{-k}((\alpha_k+2i\beta_k\cos{\pi\nu^{*}})^{*},\beta^{*}_k,\tau)&=K_{k}(\alpha_k,\beta_k,\tau)  \,.
\end{align}

\subsubsection{Spinning fields} \label{spin}
Now, we extend the cutting rules to spinning fields. We consider a very generic action for a spin $s$ field $\sigma^{i_1 i_2\dots i_s}$ as given in \cite{Bordin:2018pca},
\begin{align}
    S_{2}=\frac{1}{2s!} \int a^{3} d^{3}x dt \left[\left(\dot{\sigma}^{i_1 i_2\dots i_s}\right)^{2}-c^{2}_s a^{-2}\left(\partial _j \sigma^{i_1 i_2\dots i_s}\right)^{2}-\delta c^{2}_sa^{-2}\left(\partial_j \sigma^{j i_2\dots i_s}\right)^{2}-\left(m\sigma^{i_1 i_2\dots i_s}\right)^{2}\right] \,,
\end{align}
where $\sigma^{i_1 i_2\dots i_s}$ is a totally symmetric traceless tensor. Such an action arises in inflationary setups where the spinning field couples to the inflaton. As shown in \cite{Goodhew:2021oqg,Bordin:2018pca}, one can decompose the field in helicity basis to make the propagators diagonal i.e. different helicity components decouple (the term $\delta c^{2}_sa^{-2}\left(\partial_j \sigma^{j i_2\dots i_s}\right)^{2}$ gives the non-diagonal part). It is simplest to illustrate this for the spin-1 case. In this case the non-diagonal term in the action in Fourier space is given by,
\begin{align}
    \sigma^{i} (\vec{k})k_i k_j \sigma^{j}(-\vec{k}) \,.
\end{align}
In matrix notation (we will suppress the momentum dependence of fields), this term reads,
\begin{align}
    \pmb{\sigma}^{T}\pmb{M }\pmb{\sigma} \,,
\end{align}
where $M_{ij}=k_i k_j$. Now, we decompose the fields in terms of eigenvectors of $\pmb{M}$ i.e., 
\begin{align}
\sigma_{i} =\Sigma_{h=-1}^{+1}\sigma^{h}\epsilon^{h}_{i} \,,
\end{align}
where $\pmb{\epsilon}^{h}$ are eigenvectors of $\pmb{M}$ with eigenvalues $\lambda^{h}$. Using this fact and the condition $(\epsilon^{h})^{i}(\epsilon^{h'})^{i}=\delta ^{hh'}$, we get,
 \begin{align}
 \pmb{\sigma}^{T}\pmb{M }\pmb{\sigma}=\Sigma_{h,h'} \sigma_{h}\sigma_{h'}\lambda^{h}({\pmb{\epsilon}^{h'}})^{T}\pmb{\epsilon}^{h}=(\sigma^{h})^{2}\lambda^{h} \,,
\end{align}
 which is diagonal in helicities. Therefore, the equation of motion for each helicity component is given by,
\begin{align}
    (\sigma^{h})''-\frac{2}{\tau}(\sigma^{h})'+\left((c^{h}_s k)^{2}+m^{2}\right) \sigma^{h}=0 \,,
\end{align}
where each helicity component propagates with speed $c^{h}_s$ which is only a function of $c_s$ \& $\delta c_s$ \cite{Bordin:2018pca}. The same logic follows for spin-$s$ fields and therefore, one again gets the same equation of motion for each helicity component. The solutions to this equation are Hankel functions i.e., $H^{(1,2)}_{\nu}(-c^{h}_s k\tau)$, with $\nu=\sqrt{\frac{9}{4}-\frac{m^{2}}{H^{2}}}$. These are the same mode functions as that of massive scalars, therefore, the cutting rules are naturally extended to fields of any spin.
\subsection{Cutting rules}
\label{cutting:rule}
Given that we now have appropriate Disc operations for massive scalar fields we present the final cutting rule relation. First note that the relation between the bulk-bulk propagator and the bulk-boundary propagator is exactly as the massless and conformally coupled scalar case. The relation reads,
\begin{align}
     G_p(\tau_1,\tau_2)= iP_{p}\left[{K^{*}}_p(\tau_1)K_{p}(\tau_2)\theta(\tau_1-\tau_2)+{K^{*}}_p(\tau_2)K_{p}(\tau_1)\theta(\tau_2-\tau_1)-K_{p}(\tau_1)K_{p}(\tau_2)\right] \,.
\end{align}
Therefore, the imaginary part of a string of bulk-bulk propagators follows the same factorisation property as the one proved in \cite{Melville_2021} for the Bunch-Davies case. The only thing which changes is the definition of the cut i.e., the Disc operations. Therefore, the cutting rule reads,
\begin{align}
   \label{cutting} & i\text{Disc($m$)}\left[i\psi^{(D)}\right] \nonumber \\
 &=\underset{\text{cuts}}{\sum}\left[\prod_{\text{cut momenta}} \int P\right]\prod_{\text{sub-diagram}} (-i)\underset{\text{internal \& cut lines}}{\text{Disc($m$)}}\left[i\psi^{\left(\text{sub-diagram}\right)}\right] \,,
\end{align}
where $m$ can take values $1$ or $\tilde{2}$. In \eqref{cutting} we have suppressed the spin labels. \textit{This equation is valid for fields of any mass and spin}.

\subsection{Convergence in the far past} 
\label{convergence}
Here we address the issue of convergence of the time integrals in the far past. In \cite{Ghosh:2024aqd}, we employed an adiabatic function of the form $R(\epsilon \tau)=e^{\epsilon \tau}$ \footnote{In \cite{Ghosh:2024aqd}, it was shown that the final results are independent of the choice of the adiabatic function $R(\epsilon\tau)$.} to ensure convergence as $\tau \rightarrow -\infty$. Such a regularisation scheme is required for a Bogoliubov initial state since the mode functions include both negative- and positive-frequencies. Now, we know that the cutting rules involve wavefunction coefficients where some of the energies are taken to negative values. In the case of massless and conformally coupled scalars, one can simply replace $k$ by $-k$ since the mode functions have no branch points. However, for massive scalars due to the branch point, one needs to specify how one approaches the negative axis. In this paper, we resolved this by performing an analytic continuation to the lower half complex plane. However, note that the asymptotic limit of bulk-boundary for the Bogoliubov case has both positive and negative exponentials, therefore, in the complex plane one has,
\begin{align}
  \lim_{\tau\rightarrow-\infty}  K_k(\alpha_k,\beta_k,\tau) \sim e^{i\text{Re}(k)\tau}e^{-\text{Im}(k)\tau}+ e^{-i\text{Re}(k)\tau}e^{\text{Im}(k)\tau}\,, \hspace{10mm}\text{where} \hspace{1mm} k\in \mathbb{C}\,.
\end{align}
Naively, one might think that the integrals therefore will converge neither in the upper nor in the lower half complex plane. However, as mentioned above we always use an adiabatic function, $R(\epsilon \tau)$  to turn off interactions in the far past. This improves the convergence of the integrals. Therefore, one can perform the integrals assuming that $\epsilon>\abs{\text{Im}{(k)}}$ and then take the limit $\epsilon\rightarrow 0$. This procedure will give a convergent answer for the wavefunction coefficient.

\section{Conclusions}\label{conclusion}
In this paper, we generalised the Bogoliubov cutting rules to fields of any mass and spin. To achieve this we used generalised propagator identities, \eqref{id:1} \& \eqref{id:2} and their corresponding ``Discontinuities". Moreover, we also employed an analytic continuation of energies and consequently the wavefunction coefficients to the lower half complex plane by a careful order of limits (see Sec. \ref{convergence}). Interesting future directions include:
\begin{itemize}
    \item Deriving unitarity constraints for excited states which are not Bogoliubov transforms of the vacuum, e.g., coherent states.
    \item Understanding non-perturbative implications of unitarity for cosmological observables. If they exist, such relations can in principle be used to constrain de Sitter EFTs.
    \item Understanding implications of unitarity using the recently developed formalism of de Sitter S-matrix \cite{Melville:2023kgd,Melville:2024ove,Donath:2024utn}. It might be simpler to derive a non-perturbative relation for the S-matrix directly via $S^{\dagger}S=1$ than wavefunction or correlation functions.
    \item It is well known in the context of Flat space QFT that a violation of unitarity is tantamount to missing on-shell states in the spectrum. In light of this, a violation of cosmological unitarity relations can in principle be used to probe additional degrees of freedom during inflation.
\end{itemize}


\acknowledgments 
We would like to thank Enrico Pajer for useful discussions and valuable comments about Bunch-Davies cutting rules for massive fields. DG acknowledges support from the
Core Research Grant CRG/2023/001448 of the Anusandhan National Research Foundation
(ANRF) of the Gov. of India. 





\bibliographystyle{elsarticle-num}
\bibliography{biblio}
\end{document}